\documentclass[aapm,mph,amsmath,amssymb,reprint,hyphens]{revtex4-2}

\usepackage{graphicx}% Include figure files
\usepackage{dcolumn}% Align table columns on decimal point
\usepackage{bm}% bold math
\usepackage{xcolor}
\usepackage{url}
\usepackage{hyperref}
\usepackage[hyphenbreaks]{breakurl}

\newtheorem{theorem}{Theorem}
\newtheorem{corollary}{Corollary}[theorem]

\begin{document}

\preprint{AAPM/123-QED}

\title{Symplectic decomposition from submatrix determinants}

\author{Jason L. Pereira} \email{jason.pereira@york.ac.uk}
\affiliation{Department of Computer Science, University of York, York YO10 5GH, UK}
\affiliation{Department of Physics and Astronomy, University of Florence,
via G. Sansone 1, I-50019 Sesto Fiorentino (FI), Italy}

\author{Leonardo Banchi}
\affiliation{Department of Physics and Astronomy, University of Florence,
via G. Sansone 1, I-50019 Sesto Fiorentino (FI), Italy}
\affiliation{ INFN Sezione di Firenze, via G. Sansone 1, I-50019, Sesto Fiorentino (FI), Italy }

\author{Stefano Pirandola}
\affiliation{Department of Computer Science, University of York, York YO10 5GH, UK}

\date{\today}

\begin{abstract}
An important theorem in Gaussian quantum information tells us that we can diagonalise the covariance matrix of any Gaussian state via a symplectic transformation. Whilst the diagonal form is easy to find, the process for finding the diagonalising symplectic can be more difficult, and a common, existing method requires taking matrix powers, which can be demanding analytically. Inspired by a recently presented technique for finding the eigenvectors of a Hermitian matrix from certain submatrix eigenvalues, we derive a similar method for finding the diagonalising symplectic from certain submatrix determinants, which could prove useful in Gaussian quantum information.
\end{abstract}


\maketitle

\section{Introduction}

Gaussian states, in continuous-variable quantum information, are those states that can be described entirely by the first and second moments of their field quadratures\cite{weedbrook_gaussian_2012,ferraro2005gaussian}. In the phase-space representation, the first moments of a state can be expressed as a first moments vector and the second moments can be expressed as a covariance matrix. These are real, symmetric, positive-definite matrices. Any Gaussian unitary operation (a unitary operation that maps Gaussian states to other Gaussian states) on a state can be represented by a symplectic transformation, $S$. In other words, let $\rho$ be a Gaussian state, with covariance matrix $V$, and let $U$ be a unitary operation such that $\rho'=U\rho U^{\dagger}$ is also a Gaussian state, with covariance matrix $V'$. Then, there exists a symplectic matrix $S$ such that $V'=S^T V S$. This $S$ can therefore be regarded as the phase-space representation of $U$. Similarly, any symplectic matrix has a corresponding unitary transformation.

A key theorem, in Gaussian quantum information theory, is Williamson's theorem\cite{williamson_algebraic_1936}, which states that, for any covariance matrix, $V$, there exists some symplectic transformation, $S$, such that $V=S^T D S$, where $D$ is a covariance matrix in diagonal form. This is analogous to how every Hermitian matrix has some unitary matrix that puts it into diagonal form. The diagonalised covariance matrix is simple to find, as its non-zero elements are the eigenvalues of $|i \Omega V|$, where $\Omega$ is a matrix (defined later) called the symplectic form. The process for finding the diagonalising symplectic, $S$, is more difficult, however.

The diagonalising symplectic is required in order to find the quantum Chernoff bound\cite{audenaert_discriminating_2007} for Gaussian states\cite{pirandola_computable_2008}. This can then be used to approximate the trace norm bound on state discrimination, and has applications in quantum illumination\cite{tan_quantum_2008,karsa_quantum_2020} and quantum reading\cite{roga_device-independent_2015}, amongst other fields\cite{weedbrook_gaussian_2012}. The diagonalising symplectic is also useful for finding the optimal measurement to discriminate between Gaussian states\cite{oh_optimal_2019}.

A recent paper by Denton et al.\cite{denton_eigenvectors_2021} presented a method for finding the elements of the eigenvectors of a Hermitian matrix, using its submatrix eigenvalues. Some of the theorems presented in that paper can be simply adapted to the symplectic formalism, to provide a method for finding the diagonalising symplectic for a covariance matrix, using certain submatrix determinants. In this work, we present such a technique, along with examples of how it can be applied. We also show that our technique can be applied to certain other classes of symmetric matrix, so long as they can be diagonalised via a symplectic transformation.

\section{Finding the symplectic decomposition}\label{section: main theory}

First, we must define some quantities. Let $V$ be any valid, $d$-mode covariance matrix. $V$ is a $2d$ by $2d$, real, symmetric, square matrix. We define
\begin{equation}
    V=S^T D S,
		\label{eq: v dec}
\end{equation}
where $S$ is a symplectic matrix and $D$ is a diagonal matrix that can be expressed as
\begin{equation}
    D = \bigoplus_{m=1}^{d} \Delta_m,~~\Delta_m = \lambda_m \mathcal{I}_2.\label{eq: D form}
\end{equation}
$\mathcal{I}_n$ is the $n$ by $n$ identity matrix. $D$ is called the Williamson form of $V$ and is unique up to a rearrangement of the modes (i.e. of the labels $m$). The real values $\lambda_m$ are called the symplectic eigenvalues of $V$. The diagonalising symplectic, $S$, obeys the (symplectic) condition
\begin{equation}
    S^T \Omega S = \Omega,
		\label{eq: s cond}
\end{equation}
where $\Omega$ is called the symplectic form. $\Omega$ is required to be non-singular and skew-symmetric. Common choices in Gaussian quantum information theory are
\begin{equation}
    \Omega = \bigoplus_{m=1}^{d} \omega,~~\omega = \begin{pmatrix}
    0 &1\\
    -1 &0
    \end{pmatrix}\label{eq: symplectic form 1}
\end{equation}
and
\begin{equation}
    \Omega = \begin{pmatrix}
    0_d &\mathcal{I}_d\\
    -\mathcal{I}_d &0_d
    \end{pmatrix},\label{eq: symplectic form 2}
\end{equation}
where $0_n$ is an $n$ by $n$ square matrix with all entries equal to $0$. In this work, we will consider the symplectic form defined in Eq.~(\ref{eq: symplectic form 1}), although it is simple to convert the results to a form compatible with Eq.~(\ref{eq: symplectic form 2}). Our choice corresponds to ordering the field quadratures into the $2d$-dimensional vector $(x_1,p_1,\dots, x_d,p_d)$ where $x_j,p_j$ are, respectively, the position and momentum operators, satisfying $[x_j,p_k]=i\delta_{jk}$.

Finally we define complex numbers $s_{m,k}$ with $m=1,\dots,d$ and $k=1,\dots,2d$ such that 
\begin{align}
	\begin{split}
	&S_{2m-1,2n}= -\mathrm{Re}[s_{m,2n-1}], \\  
	&S_{2m,2n} = \mathrm{Im}[s_{m,2n-1}],  \\
	&S_{2m-1,2n-1}=\mathrm{Re}[s_{m,2n}],   \\
	&S_{2m,2n-1} =-\mathrm{Im}[s_{m,2n}].
	\end{split}
  \label{eq: Ss elements}
\end{align} 
$m=1,\dots,d$ and $n=1,\dots,d$.
Our main result is the following theorem, the proof of which is presented in Sec.~\ref{sec:proof}.

\begin{theorem}\label{theorem_main}
For any covariance matrix V, and integer indices $1\leq m\leq d$ and $1\leq k,l\leq 2d$ we may write
\begin{equation}
    \begin{split}
        \mathrm{det}\left[ R_{k,l}(V-i \lambda_m \Omega) \right] =& (-1)^{k+l} s_{m,k}^* s_{m,l}\\
        &\times \lambda_m \prod_{\substack{n=1,\\n\neq m}}^{d} (\lambda_n^2-\lambda_m^2).\label{eq: s components}
    \end{split}
\end{equation}
where $R_{k,l}(M)$, for any matrix $M$, is the matrix $M$ with the $k$-th row and $l$-th column removed, while the complex numbers $s_{k,l}$ are linearly related, via Eqs.~\eqref{eq: Ss elements}, to the real elements $S_{mn}$ of the symplectic matrix diagonalising $V$.
\end{theorem}

Based on the above theorem, we present the following algorithm for finding the diagonalising symplectic, $S$, for non-degenerate symplectic eigenvalues ($\lambda_m\neq \lambda_n$ for $n\neq m$). The degenerate case can be handled by inserting a small, $\mathcal O(\epsilon)$ perturbation that breaks the degeneracy; the correct analytical result can be obtained by taking the limit as $\epsilon\to0$, while numerical results within a certain accuracy can be obtained with a suitably small $\epsilon$ (see Section~\ref{sec:deg} for an extended discussion).
\begin{itemize}
	\item[i)]
		Find the $d$ symplectic eigenvalues, $\lambda_m$, as the positive eigenvalues of $i\Omega V$.
	\item[ii)]
	    Calculate the $d$ quantities
	    \begin{equation}
	        \aleph_m = \lambda_m \prod_{\substack{n=1,\\n\neq m}}^{d} (\lambda_n^2-\lambda_m^2),
	    \end{equation}
	    for $m=1,\dots,d$.
	\item[iii)] Fix an integer $\bar k$ between 1 and $2d$. Without loss of generality, we fix the phase such 
		that $s_{m,\bar k}$ is a positive real number -- see also the extended discussion in the next section. 
	\item[iv)]
	    For each $m=1,\dots,d$ and $l=1,\dots,2d$, calculate the $2d\times d$ minors
	    \begin{equation}
	        \beth_{\bar klm}=\mathrm{det}\left[ R_{\bar k,l}(V-i \lambda_m \Omega) \right]
	    \end{equation}
			where $\bar k$ was defined in step (iii). 
			If  $\beth_{\bar k\bar km}=0$ for some $m$, go back to step  (iii) and choose a different $\bar k$. 
		%	these are all equal to $0$, this is because $s_{m,k}=0$
%	    for any fixed $k$ (an integer between $1$ and $2d$) and $l$ taking values from $1$ to $2d$.
	\item[v)]
	    For each label $m=1,\dots,d$ and $l=1,\dots,2d$, calculate the $d\times 2d$ values 
	    \begin{equation}
				s_{m,l} = (-1)^{\bar k+l}\frac{\beth_{\bar klm}}{\sqrt{\aleph_m\beth_{\bar k\bar km}}}.
				\label{sml}
	    \end{equation}
			where again $\bar k$ was defined in step (iii). 
    \item[vi)]
        Extract the elements of $S$, using Eq.~(\ref{eq: Ss elements}).
\end{itemize}

\noindent
In Eq.~\eqref{sml} we use Eq.~\eqref{eq: s components} to write 
	    \begin{equation}
	        s_{m,\bar k}^* s_{m,l} = (-1)^{\bar k+l}\frac{\beth_{\bar klm}}{\aleph_m},
	    \end{equation}
			where $s_{m,\bar k}$ is a real positive number by explicit choice. Hence, we may calculate $s_{m,l}$ by dividing 
        the products $s_{m,k}^* s_{m,l}$ by $\sqrt{s_{m,k}^* s_{m,k}}$.

\section{Proof of the main theorem}\label{sec:proof}
In this section we prove our main result, Theorem~\ref{theorem_main}. Our proof follows similar lines to that given by Denton et al.\cite{denton_eigenvectors_2021} We begin by proving a theorem similar to their Lemma~13 and then use it to get an identity similar to their Theorem~1 and its generalisation in Proposition~17 (Lemmas 1 and 2 respectively in the first version preprint on arXiv).

Note the following equations, which follow directly from the previous definitions:
\begin{align}
    &\Omega^{-1}=\Omega^{T}=-\Omega,\\
    &S^{-1}=-\Omega S^T \Omega,\label{eq: S inverse}\\
    &D = -S \Omega V \Omega S^T.\label{eq: D from V}
\end{align}
Next, we define the matrices $A_m$, which are given by
\begin{equation}
    A_m = V - i \lambda_m \Omega.
\end{equation}
Note that
\begin{equation}
    -S \Omega A_m \Omega S^T = D - i \lambda_m \Omega, 
\end{equation}
so $-S \Omega A_m \Omega S^T$ is a block diagonal, Hermitian matrix, composed of two by two blocks along the main diagonal.
 Note too that the symplectic eigenvalues of $V$ can be easily found as the eigenvalues of $|i \Omega V|$.

Let $U$ be the $2d$ by $2d$ matrix with elements
\begin{equation}
    U_{m,n}=\frac{1}{\sqrt{2}}\begin{cases}
    \delta_{2m-1,n} + i \delta_{2m,n} &m \leq d\\
    \delta_{2(m-d)-1,n} - i \delta_{2(m-d),n} &m > d
    \end{cases}.\label{eq: U form}
\end{equation}
Note that $U$ is a unitary. To make its structure clearer, its explicit form for $d=3$ is
\begin{equation}
    U_{d=3}=\frac{1}{\sqrt{2}}\begin{pmatrix}
    1 &i &0 &0 &0 &0\\
    0 &0 &1 &i &0 &0\\
    0 &0 &0 &0 &1 &i\\
    1 &-i &0 &0 &0 &0\\
    0 &0 &1 &-i &0 &0\\
    0 &0 &0 &0 &1 &-i
    \end{pmatrix}.
\end{equation}
The matrix $U$ maps the vector of quadrature operators to the vector of creation and annihilation  operators $(a_1,\dots,a_d,a_1^\dagger,\dots,a_d^\dagger)$, where $a_j = (x_j+ip_j)/\sqrt 2$. 
$U$ diagonalises the matrices $D - i \lambda_m \Omega$. We can write
\begin{equation}
    D'_m = U(D - i \lambda_m \Omega)U^{\dagger}=-U S \Omega A_m \Omega S^T U^{\dagger},\label{eq: diagonalising A}
\end{equation}
where $D'_m$ is a diagonal matrix, with its elements along the main diagonal given by
\begin{equation}
    (D'_m)_{n,n}=\begin{cases}
    \lambda_n - \lambda_m &n \leq d\\
    \lambda_n + \lambda_m &n > d
    \end{cases}.\label{eq: D' elements}
\end{equation}
Note that the $m$-th element on the main diagonal (i.e. $(D'_m)_{m,m}$) will always be $0$.

We now define the $2d$ by $1$ column vectors $e_m$, which have elements $(e_m)_n=\delta_{m,n}$, where $\delta$ is the Kronecker delta function and the label $m$ runs from $1$ to $d$. Note that the label $m$ could run up to $2d$, but we do not use vectors $e_{m>d}$. We can then define the vectors $s_m$ as
\begin{equation}
    s_m = -\sqrt{2}\Omega S^T U^{\dagger} e_m.\label{eq: s_m}
\end{equation}
These $d$ vectors contain all of the elements of $S$. To show this, let us carry out the multiplication in Eq.~(\ref{eq: s_m}) step by step. $U^{\dagger}e_m$ is a column vector with elements
\begin{equation}
    (U^{\dagger}e_m)_n = \frac{1}{\sqrt{2}} (\delta_{2m-1,n}-i\delta_{2m,n}).
\end{equation}
$S^T U^{\dagger}e_m$ is a column vector with elements
\begin{equation}
    (S^T U^{\dagger}e_m)_n = \frac{1}{\sqrt{2}} (S_{2m-1,n}-i S_{2m,n}).
\end{equation}
Finally, $s_m$ is the $2d$ by $1$ column vector with elements
\begin{equation}
    (s_m)_n=\begin{cases}
    -S_{2m-1,n+1}+i S_{2m,n+1} &\mathrm{odd}~n\\
    S_{2m-1,n-1}-i S_{2m,n-1} &\mathrm{even}~n
    \end{cases}.\label{eq: s elements}
\end{equation}
We will sometimes refer to the $n$-th element of $s_m$ as $s_{m,n}$. The vectors $s_m$ can be regarded as symplectic counterparts to eigenvectors, in the same way that the symplectic eigenvalues of a covariance matrix are to eigenvalues.

The form of $U$ given in Eq.~(\ref{eq: U form}) is the only expression given so far that explicitly depends on the choice of symplectic form, $\Omega$. If we want to use the symplectic form given by Eq.~(\ref{eq: symplectic form 2}), instead of the form given by Eq.~(\ref{eq: symplectic form 1}), we can use the same expressions, but with the definition of the vectors $s_m$ changed so that they have elements
\begin{equation}
    (s_m)_n=\begin{cases}
    -S_{m,n+d}+i S_{m+d,n+d} &n \leq d\\
    S_{m,n-d}-i S_{m+d,n-d} &n > d
    \end{cases}.
\end{equation}
With the $s_m$ vectors redefined in this way, all of the theorems given in this paper hold for the symplectic form given by Eq.~(\ref{eq: symplectic form 2}).

We are now ready to state Theorem~\ref{prop: det equality}.

\begin{theorem}\label{prop: det equality}
For any pair of $2d$ by $2d-1$ matrices, $B_x$ and $B_y$,
\begin{equation}
    \begin{split}
        \mathrm{det}\left[B_x^{\dagger}A_m B_y\right]=&\mathrm{det}\left[ \left(B_x~\frac{s_m}{\sqrt{2}}\right) \right]^{*}\mathrm{det}\left[ \left(B_y~\frac{s_m}{\sqrt{2}}\right) \right]\\
        &\times 2 \lambda_m \prod_{\substack{n=1,\\n\neq m}}^{d} (\lambda_n^2-\lambda_m^2),
    \end{split}\label{eq: det equality}
\end{equation}
where $\left(B_{x/y}~\frac{s_m}{\sqrt{2}}\right)$, on the right-hand side of the equation, indicates concatenation of $B_{x/y}$ and $\frac{s_m}{\sqrt{2}}$ into a single $2d$ by $2d$, square matrix, rather than multiplication, and $*$ indicates the complex conjugate.
\end{theorem}

The theorem holds for all matrices $B_x$ and $B_y$ (whether real or complex) and for all covariance matrices $V$, although both sides of Eq.~(\ref{eq: det equality}) will go to $0$ if $\lambda_m$ is a degenerate symplectic eigenvalue.

To prove this theorem, let us define
\begin{equation}
    B'_{x/y} = U \Omega S B_{x/y},
\end{equation}
so that we can express $B_{x/y}$ as
\begin{equation}
    B_{x/y} = -\Omega S^T U^{\dagger} B'_{x/y}.\label{eq: B in terms of B'}
\end{equation}
Substituting Eq.~(\ref{eq: B in terms of B'}) into $B^{\dagger}A_m B$, we get
\begin{equation}
    \begin{split}
        B_x^{\dagger}A_m B_y &= (B_x'^{\dagger} U S \Omega)A_m (-\Omega S^T U^{\dagger} B_y')\\
        &= B_x'^{\dagger} D'_m B_y,
    \end{split}
\end{equation}
where we have used Eq.~(\ref{eq: diagonalising A}).

Next we define $R_{m,n}(M)$, for any matrix $M$, as the matrix $M$ with the $m$-th row and the $n$-th column removed. The index $0$ indicates that no rows/columns are removed (e.g. $R_{0,m}(M)$ is the matrix $M$ with the $m$-th column removed but no rows removed). Recalling that $D'_m$ is a diagonal matrix with the $m$-th element on the main diagonal equal to $0$, we can write
\begin{equation}
    \begin{split}
        \mathrm{det}\left[B_x'^{\dagger} D'_m B_y'\right] =& \mathrm{det}\left[R_{m,0}(B_x')^{\dagger} R_{m,m}(D') R_{m,0}(B_y')\right]\\
        =& \mathrm{det}\left[R_{m,0}(B'_x) \right]^*\mathrm{det}\left[R_{m,0}(B'_y) \right]\\
        &\times \mathrm{det}\left[R_{m,m}(D')\right],
    \end{split}
\end{equation}
where, on the second line, we have used the cyclic invariance of the determinant of a product of square matrices. Recalling Eq.~(\ref{eq: D' elements}), we can write
\begin{equation}
    \begin{split}
        \mathrm{det}\left[R_{m,m}(D')\right] &= \prod_{\substack{n=1,\\n\neq m}}^{d} (\lambda_n-\lambda_m) \prod_{n=1}^{d} (\lambda_n+\lambda_m)\\
        &= 2\lambda_m \prod_{\substack{n=1,\\n\neq m}}^{d} (\lambda_n^2-\lambda_m^2).
    \end{split}
\end{equation}
Thus, the left-hand side of Eq.~(\ref{eq: det equality}) is given by
\begin{equation}
    \begin{split}
        \mathrm{det}\left[B_x^{\dagger}A_m B_y\right] =& \mathrm{det}\left[R_{m,0}(B'_x) \right]^*\mathrm{det}\left[R_{m,0}(B'_y) \right]\\
        &\times 2\lambda_m \prod_{\substack{n=1,\\n\neq m}}^{d} (\lambda_n^2-\lambda_m^2).
    \end{split}
\end{equation}

Finally, we must compute the right-hand side of Eq.~(\ref{eq: det equality}). Using Eqs.~(\ref{eq: s_m}) and (\ref{eq: B in terms of B'}), we can write
\begin{equation}
    \begin{split}
        \mathrm{det}\left[ \left(B_{x/y}~\frac{s_m}{\sqrt{2}}\right) \right] &= \mathrm{det}\left[-\Omega S^T U^{\dagger} (B_{x/y}'~e_m) \right]\\
        &= \mathrm{det}\left[-\Omega S^T U^{\dagger}\right] \mathrm{det}\left[ (B_{x/y}'~e_m) \right].
    \end{split}
\end{equation}
The determinants of $\Omega$ and $S$ are both $1$, so we can write
\begin{equation}
    \mathrm{det}\left[ \left(B_{x/y}~\frac{s_m}{\sqrt{2}}\right) \right] = -\mathrm{det}\left[ U^{\dagger} \right]\mathrm{det}\left[ (B_{x/y}'~e_m) \right],
\end{equation}
where the determinant of $U^{\dagger}$ also has a magnitude of 1.

For any lower-triangular block matrix $M$, expressible as
\begin{equation}
    M = \begin{pmatrix}
    X &0\\
    Y &Z
    \end{pmatrix},\label{eq: lower triangular block matrix}
\end{equation}
where the $0$ represents a matrix composed entirely of $0$s, we know that
\begin{equation}
    \mathrm{det}[M] = \mathrm{det}[X]\mathrm{det}[Z].\label{eq: block matrix det}
\end{equation}
Using the form of $e_m$, the matrix $(B'_{x/y}~e_m)$ can be expressed in the form given by Eq.~(\ref{eq: lower triangular block matrix}), where $Z$ is the single element $1$, up to some permutation of the rows (in fact, a single swap). Recalling that a permutation of the rows of a matrix  simply multiplies the determinant by $\pm 1$, we can write
\begin{equation}
    \begin{split}
        \mathrm{det}\left[ \left(B_{x/y}~\frac{s_m}{\sqrt{2}}\right) \right] =& (-1)^p\mathrm{det}\left[ U^{\dagger} \right]\\
        &\times\mathrm{det}\left[R_{m,0}(B_{x/y}')\right],
    \end{split}
\end{equation}
where $p$, which depends on the parity of the permutation, is an integer. Note that
\begin{equation}
    \mathrm{det}\left[ U^{\dagger} \right]^* = \mathrm{det}\left[ U \right] = \mathrm{det}\left[ U^{\dagger} \right]^{-1},
\end{equation}
and so
\begin{equation}
    \begin{split}
        \mathrm{det}\left[ \left(B_{x}~\frac{s_m}{\sqrt{2}}\right) \right]^* =& (-1)^p\mathrm{det}\left[ U^{\dagger} \right]^{-1}\\
        &\times\mathrm{det}\left[R_{m,0}(B_{x}')\right]^*.
    \end{split}
\end{equation}
Since the left-hand side of Eq.~(\ref{eq: det equality}) equals the right-hand side, this concludes the proof of Theorem~\ref{prop: det equality}.

If $\lambda_m$ is a degenerate symplectic eigenvalue, $D'_m$ will have multiple elements on the main diagonal that are equal to 0, so both sides of Eq.~(\ref{eq: det equality}) will trivially go to $0$. A generalisation of Theorem~\ref{prop: det equality} is given by Corollary~\ref{corollary theorem 1 degenerate}.

\begin{corollary}\label{corollary theorem 1 degenerate}
Let $\lambda_m$ be an $k$-fold degenerate symplectic eigenvalue of $V$, and let the labels for the degenerate eigenvalues form the set $\{m_j\}$, where $j$ runs from $1$ to $k$. For any pair of $2d$ by $2d-k$ matrices, $B_x$ and $B_y$,
\begin{equation}
    \begin{split}
        \mathrm{det}\left[B_x^{\dagger}A_m B_y\right]=&\mathrm{det}\left[ \left(B_x~\frac{s_{m_1}}{\sqrt{2}}\dots \frac{s_{m_k}}{\sqrt{2}}\right) \right]^*\\
        &\times \mathrm{det}\left[ \left(B_y~\frac{s_{m_1}}{\sqrt{2}}\dots \frac{s_{m_k}}{\sqrt{2}}\right) \right]\\
        &\times (2\lambda_m)^k \prod_{\substack{n=1,\\n\notin \{m_j\}}}^{d} (\lambda_n^2-\lambda_m^2),
    \end{split}
\end{equation}
where $\left(B_{x/y}~\frac{s_{m_1}}{\sqrt{2}}\dots \frac{s_{m_k}}{\sqrt{2}}\right)$, on the right-hand side of the equation, is the concatenation of $B_{x/y}$ with all $k$ of the vectors in $\{\frac{s_{m_j}}{\sqrt{2}}\}$, to form a $2d$ by $2d$, square matrix.
\end{corollary}

Corollary~\ref{corollary theorem 1 degenerate} can be proved in exactly the same way as Theorem~\ref{prop: det equality}, but using the new definition of $B_{x/y}$ and replacing $R_{m,0}(B'_{x/y})$ and $R_{m,m}(D')$  with $R_{\{m_j\},0}(B'_{x/y})$ and $R_{\{m_j\},\{m_j\}}(D')$ respectively (where all of the rows/columns labelled by elements in the subscripted sets are removed).

We can now prove the relationship between the elements of the diagonalising symplectic and the submatrix determinants of the matrices $A_m$ given in Theorem~\ref{theorem_main}, which we restate here for clarity.

\setcounter{theorem}{0}
\begin{theorem}
For any covariance matrix V,
\begin{equation}
    \begin{split}
        \mathrm{det}\left[ R_{k,l}(V-i \lambda_m \Omega) \right] =& (-1)^{k+l} s_{m,k}^* s_{m,l}\\
        &\times \lambda_m \prod_{\substack{n=1,\\n\neq m}}^{d} (\lambda_n^2-\lambda_m^2).
    \end{split}
\end{equation}
\end{theorem}

Define $M_1$ as the concatenation of a $1$ by $2d-1$ row vector with all of its elements equal to $0$ and a $(2d-1)$-dimensional identity matrix, i.e.
\begin{equation}
    M_1 = \begin{pmatrix}
    0\\
    \mathcal{I}_{2d-1}
    \end{pmatrix}.
\end{equation}
Then define the matrices $M_l$ as $M_1$ with the rows permuted such that the row vector of $0$s is the $l$-th row.

Theorem~\ref{theorem_main} now follows directly from setting $B_x$, in Theorem~\ref{prop: det equality}, to $M_k$ and setting $B_y$ to $M_l$. For the left-hand side, we use
\begin{equation}
    M_l^{\dagger}AM_k = R_{k,l}(A),
\end{equation}
and for the right-hand side we use Eq.~(\ref{eq: block matrix det}) (albeit with an upper-triangular block matrix, rather than a lower-triangular one). The factor of $\pm 1$ comes from the parity of the permutation of rows required to put the matrix in an upper-triangular form.

If we fix the phase of one element of $s_m$, we now have an expression for the magnitudes and relative phases of the other elements of $s_m$, except for cases in which $\lambda_m$ is a degenerate symplectic eigenvalue. The only remaining task is then to fix the relative phases of the different $s_m$. In fact, this is not required, because these relative phases are arbitrary. There is a correspondence here with the fact that, when finding the diagonalising unitary for a Hermitian matrix, the relative phases between the eigenvectors are also arbitrary.

Let $S$ be the symplectic recovered using Theorem~\ref{theorem_main} with some arbitrary set of relative phases between the vectors $s_m$ and let $S'$ be the symplectic corresponding to some different set of relative phases, so that
\begin{equation}
    s'_m=e^{i\phi_m}s_m.
\end{equation}
From Eq.~(\ref{eq: s elements}), we can see that
\begin{align}
    &S'=PS,\\
    &P=\bigoplus_{m=1}^d \begin{pmatrix}
    \cos \phi_m &\sin \phi_m\\
    -\sin \phi_m &\cos \phi_m
    \end{pmatrix}.\label{eq: 1 mode rots}
\end{align}
$P$ is a symplectic corresponding to a series of one-mode phase rotations, so $S'$ is also a valid symplectic matrix. Further, suppose $S$ diagonalises a covariance matrix, $V$. $S'$ will then also diagonalise $V$, because the Williamson form is invariant under the application of one-mode phase rotations. Thus, by finding the $d$ symplectic eigenvalues and $2d^2$ submatrix determinants, we can completely determine the diagonalising symplectic.

The fact that the matrix we find obeys the symplectic condition is implicit to the method. We start by assuming that such a matrix exists (which is guaranteed by Williamson's theorem\cite{williamson_algebraic_1936}) and then calculate the values of its elements. Then, since all matrices that obey Eq.~(\ref{eq: s components}) are related by symplectic transformations of the form in Eq.~(\ref{eq: 1 mode rots}), the matrix found is guaranteed to be a diagonalising symplectic.

For completeness, we can also apply Corollary~\ref{corollary theorem 1 degenerate} to find a corollary to Theorem~\ref{theorem_main} that applies in the degenerate case.

\begin{corollary}\label{corollary theorem 2 degenerate}
Let $\lambda_m$ be an $p$-fold degenerate symplectic eigenvalue of $V$, and let the labels for the degenerate eigenvalues form the $p$-element set $\{m_j\}$. Let $\{k_j\}$ and $\{l_j\}$ be two other sets, each consisting of $p$ elements drawn from the set of integers between $1$ and $2d$.
\begin{equation}
    \begin{split}
        \mathrm{det}\left[ R_{\{k_j\},\{l_j\}}(V-i \lambda_m \Omega) \right]=& (-1)^{\sum k_j+l_j} \lambda_m^p\\
        &\times \mathrm{det}\left[ s_{\{m_j\},\{k_j\}} \right]^*\\
        &\times \mathrm{det}\left[ s_{\{m_j\},\{l_j\}} \right]\\
        &\times \prod_{\substack{n=1,\\n\notin \{m_j\}}}^{d} (\lambda_n^2-\lambda_m^2),
    \end{split}
\end{equation}
where $s_{\{m_j\},\{k/l_j\}}$ denotes the $p$ by $p$, square matrix whose elements are $s_{x,y}$, where $x$ is drawn from $\{m_j\}$ and $y$ is drawn from $\{k/l_j\}$.
\end{corollary}

Note the correspondence with Proposition~17 from Denton et al\cite{denton_eigenvectors_2021}.

Corollary~\ref{corollary theorem 2 degenerate} gives us $\binom{2d}{p}$ conditions (since this is the number of different choices of sets $\{k/l_j\}$) on the $2dp$ unknown elements of $\{s_{m_j}\}$.

\section{Handling the degenerate case}\label{sec:deg}

Theorem~\ref{theorem_main} holds for all covariance matrices, but does not always yield useful results. If the covariance matrix, $V$, has degenerate symplectic eigenvalues, $\{\lambda_{m_j}\}$, we will not be able to use Eq.~(\ref{eq: s components}) to find $\{s_{m_j}\}$, as both sides of the equation will go to $0$.

One workaround is to introduce a perturbed matrix, $V'$. Specifically, we define
\begin{equation}
    V'=V+\epsilon\Delta,
\end{equation}
where $\Delta$ is some perturbation matrix, chosen such that $V'$ is a valid covariance matrix with non-degenerate symplectic eigenvalues for $0 < \epsilon < E$, where $E$ is some positive, real number. In other words, we can treat $V$ as the limiting case of some sequence of covariance matrices, parametrised by $\epsilon$, that do not generally have degenerate eigenvalues. We then find the diagonalising symplectic for $V'$, parametrised by $\epsilon$, and then take the limit of our expression for this diagonalising symplectic as $\epsilon\to 0$. This technique is demonstrated by the example in Section~\ref{section: degenerate example}.

In order to prove that this technique is valid in all cases, we must show that the elements of the diagonalising symplectic are continuous functions of the covariance matrix. First, let us clarify what this means. Let $f$ be a function mapping from a variable $\epsilon$ to the set of complex numbers. Restrict epsilon to the real interval $\mathcal{D}=[0,E]$. We call $f$ continuous at the point $\epsilon_0$ iff, for any $\delta$ there exists some $\delta_{\epsilon}$ such that
\begin{equation}
    \left|f[\epsilon_0] - f[\epsilon_1]\right|<\delta~~\forall \epsilon_1 \in \mathcal{D}:|\epsilon_0-\epsilon_1|<\delta_{\epsilon}.
\end{equation}
In other words, $f$ is continuous at $\epsilon_0$ iff there is some neighbourhood around $\epsilon_0$ within which all values of $\epsilon$ give rise to a value of $f$ that is sufficiently close to $f[\epsilon_0]$. If all elements of a matrix are continuous functions of $\epsilon$, we will say that that matrix is a continuous function of $\epsilon$.

An important ingredient in our proof is the fact that the eigenvalues of a matrix parameterised by a variable restricted to a real interval are continuous functions of that variable\cite{kato_perturbation_1995,li_eigenvalue_2019}. This stems from the fact that the eigenvalues of a matrix are the solutions of that matrix's characteristic equation, which is a polynomial in its elements.

Recall that the symplectic eigenvalues of a covariance matrix $V$ are the absolute values of the eigenvalues of $i\Omega V$ (a Hermitian matrix), and that they are non-zero for all valid covariance matrices (in fact, they are greater than or equal to the shot-noise, the numerical value of which depends on the convention used\cite{weedbrook_gaussian_2012}). Consequently, the symplectic eigenvalues of $V'$, $\lambda'_m$, are continuous functions of $\epsilon$ in the domain $\mathcal{D}$. It is immediate that products and sums of continuous functions are themselves continuous, so
\begin{equation}
    \aleph_m=\lambda'_m \prod_{\substack{n=1,\\n\notin \{m_j\}}}^{d} (\lambda_n^{\prime 2}-\lambda_m^{\prime 2})
\end{equation}
is a continuous function of $\epsilon$.

Since $R_{k,l}(V'-i \lambda'_m \Omega)$ is a continuous function of $\epsilon$, its eigenvalues are too, and so $\mathrm{det}\left[ R_{\{k_j\},\{l_j\}}(V-i \lambda_m \Omega) \right]$ is also a continuous function of $\epsilon$.

Putting this together, we can see that the products $s^{\prime *}_{m,k}s^{\prime}_{m,l}$ and hence the elements $s'_{m,l}$ are continuous functions of $\epsilon$. Finally, since the real/imaginary part of the output of a continuous function is also continuous, the calculated symplectic, $S'(\epsilon)$, is a continuous function of $\epsilon$ in the domain $\mathcal{D}$.

We are guaranteed that there exists some matrix $\Delta$ such that $V'$ is a valid covariance matrix with non-degenerate symplectic eigenvalues for sufficiently small, positive $\epsilon$. For instance, we could choose
\begin{equation}
    \Delta = S^T \xi S,\label{eq: possible peturbation matrix}
\end{equation}
where $\xi$ is some diagonal matrix. Then, after diagonalisation, the Williamson form of the perturbed matrix becomes
\begin{equation}
    D'=D+\epsilon\xi.
\end{equation}
We can therefore always choose $\xi$ such that $V'$ is non-degenerate. Note that $S$ is not known (indeed the purpose of the algorithm is to find it), so Eq.~(\ref{eq: possible peturbation matrix}) is not intended to provide a viable choice of $\Delta$ for the method, but rather to demonstrate that such a matrix is always guaranteed to exist.

\section{Extension to indefinite and non-real, symmetric matrices}
A symmetric matrix can be put in a positive, diagonal form via a real, symplectic transformation if and only if it is real and positive. However, there exist indefinite and non-real, symmetric matrices that can be put in a diagonal form via a real, symplectic transformation (although this diagonal form will not be positive-definite).

To understand this, let us first expand the definition of symplectic eigenvalues slightly. Let $V$ be a symmetric matrix. Suppose there exists some real, symplectic matrix $S$ such that $V=S^T D S$, where $D$ can be expressed as in Eq.~(\ref{eq: D form}), but without any condition on the $\lambda_m$ (i.e. they may now be negative or complex instead of all being real and positive). Note that this is not guaranteed to be the case, and such a transformation may not exist at all. It is clear that we can no longer determine the diagonal form solely by taking the absolute values of the eigenvalues of $i\Omega V$, since these will always be positive values. However, it remains true that the eigenvalues of $i\Omega V$ are invariant under symplectic transformations. We can show this by recalling that these eigenvalues are determined by the characteristic equation
\begin{equation}
    \mathrm{det}[i\Omega V - \lambda \mathcal{I}_d]=0.
\end{equation}
Then, since the determinant of a product of square matrices is the product of the determinants of the matrices, we can write
\begin{equation}
    \mathrm{det}[i\Omega V - \lambda \mathcal{I}_d]=\mathrm{det}[i S^{-1}\Omega V S - \lambda S^{-1}S],
\end{equation}
and finally
\begin{equation}
    \mathrm{det}[i\Omega V - \lambda \mathcal{I}_d]=\mathrm{det}[i\Omega S^T V S - \lambda \mathcal{I}_d],
\end{equation}
where we have applied Eq.~(\ref{eq: S inverse}). The eigenvalues of $i\Omega D$, where $D$ is given by Eq.~(\ref{eq: D form}), are $\pm \lambda_m$. The difficulty now comes from determining which of these eigenvalues are $\lambda_m$ and which are $-\lambda_m$. This was not an issue in the case of positive, symmetric matrices, since then we knew $\lambda_m>0$. Henceforth, we shall refer to the values with positive real parts as the positive symplectic eigenvalues, $\lambda_m^{+}$, and the values with negative real parts as the negative symplectic eigenvalues, $\lambda_m^{-}$. If the real part is 0, we can instead assign the eigenvalue with a positive imaginary part as the positive symplectic eigenvalue.

Our proofs hold regardless of the values of the symplectic eigenvalues, so if we know the symplectic eigenvalues (including their signs), we can still calculate the diagonalising symplectic via the method we have presented. However, suppose we do not know the diagonal form in advance, only that the matrix $V$ can be diagonalised via some real, symplectic transformation. How then can we determine whether $\lambda_m^{+}$ or $\lambda_m^{-}$ appears in the diagonal form, $D$?

First, consider the effect on Eq.~(\ref{eq: s components}) of substituting $-\lambda_m$ for $\lambda_m$. Specifically, let us consider the case in which $k=l$. The left-hand side of the equation is unchanged, because $R_{k,k}(V-i \lambda_m \Omega)=R^T_{k,k}(V+i \lambda_m \Omega)$. The right-hand sign picks up a minus sign (note that the signs of the other symplectic eigenvalues do not matter, since they are squared). Now, suppose we use Eq.~(\ref{eq: s components}) to calculate $s_{m,k}^{*} s_{m,k}$. Since this quantity is equal to $|s_{m,k}|^2$, it should always be a positive quantity. If we have used the ``wrong" symplectic eigenvalue (i.e. used $\lambda_m^{+}$ when $\lambda_m=\lambda_m^{-}$, or vice versa), we will get a negative quantity (specifically, minus the correct value). The sign of the quantity
\begin{equation}
    \gimel_m = \frac{\mathrm{det}[R_{k,k}(V-i \lambda_m^{+} \Omega)]}{\lambda_m^{+} \prod_{\substack{n=1,\\n\neq m}}^{d} (\lambda_n^2-\lambda_m^{+2})},
\end{equation}
where $k$ can be any integer between $1$ and $2d$, determines whether $\lambda_m^{+}$ or $\lambda_m^{-}$ appears in the diagonal form. Specifically, if the sign is positive, $\lambda_m=\lambda_m^{+}$ and if it is negative, $\lambda_m=\lambda_m^{-}$. We therefore now have a way to calculate the symplectic eigenvalues, including the signs, for symmetric matrices that are not both real and positive.

We can therefore modify our algorithm to find the diagonalising symplectic in the following way. In step (i), the symplectic eigenvalues are instead calculated as the eigenvalues of $i\Omega V$ that have positive real parts (i.e. we choose the positive symplectic eigenvalues). In step (iv), we first calculate the $d$ values $\beth_{\bar{k}\bar{k}m}$ (i.e. before calculating the values for which $l\neq\bar{k}$). Then, if the sign of
\begin{equation}
    \gimel_m = \frac{\beth_{\bar{k}\bar{k}m}}{\aleph_m}
\end{equation}
is negative, replace $\aleph_m$ with $-\aleph_m$ and $\lambda_m$ with $-\lambda_m$ and then calculate all of the remaining $\beth_{\bar{k}lm}$ values using the new value of $\lambda_m$. The rest of the steps are then unchanged.

It should be noted that the method does not work if one of the symplectic eigenvalues is $0$. This is because - as in the case in which $V$ has degenerate symplectic eigenvalues - both sides of Eq.~(\ref{eq: s components}) go to $0$. We can resolve this issue by, again, inserting a small perturbation to slightly change the symplectic eigenvalues.

Note too that this method yields the diagonalising symplectic if and only if such a symplectic exists, which is no longer guaranteed if the symmetric matrix is not real and positive-definite. In fact, since Theorem~\ref{theorem_main} holds for any symmetric matrix that can be diagonalised by a real, symplectic transformation (into a matrix of the form given in Eq.~(\ref{eq: D form})), it implicitly provides a condition for such a diagonalising symplectic to exist. If the technique presented here does not yield a valid, diagonalising, symplectic matrix (and none of the symplectic eigenvalues are degenerate or equal to $0$), then such a matrix does not exist. By extension, this means that $\mathrm{Im}[\gimel_m]=0$ for all $m$ is a necessary condition for a symmetric matrix to be diagonalisable. One final thing to note is that we could consider other diagonal forms that do not obey Eq.~(\ref{eq: D form}), e.g. a form in which each $2$ by $2$ block can be written as $\mathrm{diag}(\lambda_m,\lambda_m^*)$ rather than $\mathrm{diag}(\lambda_m,\lambda_m)$, but that is beyond the scope of this work.

\section{Examples}
Let us consider some examples of how Theorem~\ref{theorem_main} can be applied.

\subsection{Two-mode squeezed states}
The symplectic matrix that diagonalises a two-mode squeezed state is known~\cite{SiHuiGaussianQI,weedbrook_gaussian_2012}, and so provides a useful example case. The covariance matrices we are interested in take the form
\begin{equation}
    V=\begin{pmatrix}
    a \mathcal{I}_2 &c \mathcal{Z}\\
    c \mathcal{Z} &b \mathcal{I}_2
    \end{pmatrix},\label{eq: CM 2-mode squeezed}
\end{equation}
where $\mathcal{Z}$ is the Pauli-Z matrix. The symplectic eigenvalues of $V$ are
\begin{align}
    &\lambda_{1}=\frac{\sqrt{y}-(a-b)}{2},~~\lambda_{2}=\frac{\sqrt{y}+(a-b)}{2},\\
    &y=(a+b)^2-4c^2.
\end{align}

Now define
\begin{align}
    &\aleph_1=\lambda_1 (\lambda_2^2-\lambda_1^2),\\
    &\aleph_2=\lambda_2 (\lambda_1^2-\lambda_2^2).
\end{align}
We can then calculate
\begin{align}
    &\aleph_1=\frac{1}{2}\left((a-b)y - (a-b)^2\sqrt{y}\right),\label{eq: 2 mode example d_1}\\
    &\aleph_2=-\frac{1}{2}\left((a-b)y + (a-b)^2\sqrt{y}\right).\label{eq: 2 mode example d_2}
\end{align}

Next, we calculate four different submatrix determinants (minors) of the matrix
\begin{equation}
    V - i \lambda_1 \Omega = \begin{pmatrix}
    a &-i \lambda_1 &c &0\\
    i \lambda_1 &a &0 &-c\\
    c &0 &b &-i \lambda_1\\
    0 &-c &i \lambda_1 &b
    \end{pmatrix}.
\end{equation}
Specifically, we find the minors
\begin{equation}
    \beth_{2,j,1}=\mathrm{det}\left[ R_{2,j}(V-i \lambda_1 \Omega) \right],
\end{equation}
for $j$ taking values from $1$ to $4$. We calculate
\begin{align}
    &\beth_{2,1,1}=i \lambda_1 (\lambda_1^2-b^2+c^2),\label{eq: 2 mode example det 1}\\
    &\beth_{2,2,1}=a(b^2-\lambda_1^2)-bc^2,\\
    &\beth_{2,3,1}=-i \lambda_1 c(a-b),\\
    &\beth_{2,4,1}=-c(\lambda_1^2-ab+c^2)\label{eq: 2 mode example det 4}
\end{align}
Combining Eq.~(\ref{eq: 2 mode example d_1}) with Eqs.~(\ref{eq: 2 mode example det 1}) to (\ref{eq: 2 mode example det 4}), and using Theorem~\ref{theorem_main}, we get
\begin{align}
    &s_{1,2}^* s_{1,1} = -i\frac{a+b-\sqrt{y}}{2\sqrt{y}},\\
    &s_{1,2}^* s_{1,2} = \frac{a+b-\sqrt{y}}{2\sqrt{y}},\\
    &s_{1,2}^* s_{1,3} = i\frac{c}{\sqrt{y}},\\
    &s_{1,2}^* s_{1,4} = \frac{c}{\sqrt{y}}.
\end{align}
Finally, we calculate
\begin{align}
    &s_1=\begin{pmatrix}
    -i \omega_- &\omega_- &i \omega_+ &\omega_+
    \end{pmatrix},\\
    &\omega_{\pm}=\sqrt{\frac{a+b\pm\sqrt{y}}{2\sqrt{y}}}.
\end{align}

Similarly, we find the minors
\begin{equation}
    \beth_{4,j,2}=\mathrm{det}\left[ R_{4,j}(V-i \lambda_2 \Omega) \right],
\end{equation}
for $j$ taking values from $1$ to $4$, by calculating
\begin{align}
    &\beth_{4,1,2}=i \lambda_2 c (a-b),\label{eq: 2 mode example det 5}\\
    &\beth_{4,2,2}=-c(\lambda_2^2-ab+c^2),\\
    &\beth_{4,3,2}=i \lambda_2 (\lambda_2^2-a^2+c^2),\\
    &\beth_{4,4,2}=b(a^2-\lambda_2^2)-ac^2.\label{eq: 2 mode example det 8}
\end{align}
Combining Eq.~(\ref{eq: 2 mode example d_2}) with Eqs.~(\ref{eq: 2 mode example det 5}) to (\ref{eq: 2 mode example det 8}), we get
\begin{align}
    &s_{2,4}^* s_{2,1} = i\frac{c}{\sqrt{y}},\\
    &s_{2,4}^* s_{2,2} = \frac{c}{\sqrt{y}},\\
    &s_{2,4}^* s_{2,3} = -i\frac{a+b-\sqrt{y}}{2\sqrt{y}},\\
    &s_{2,4}^* s_{2,4} = \frac{a+b-\sqrt{y}}{2\sqrt{y}}.
\end{align}
Finally, we can calculate
\begin{equation}
    s_2=\begin{pmatrix}
    i \omega_+ &\omega_+ &-i \omega_- &\omega_-
    \end{pmatrix}.
\end{equation}

Using Eq.~(\ref{eq: s elements}), we can construct the diagonalising symplectic, $S$, from the elements of $s_1$ and $s_2$. We get
\begin{equation}
    S=\begin{pmatrix}
    \omega_- \mathcal{Z} &\omega_+ \mathcal{I}_2\\
    \omega_+ \mathcal{I}_2 &\omega_- \mathcal{Z}
    \end{pmatrix}.
\end{equation}
This is the same as the expression given by Ref.~[\onlinecite{SiHuiGaussianQI}] up to multiplication by
\begin{equation}
    U=\begin{pmatrix}
    0_2 &\mathcal{I}_2\\
    \mathcal{I}_2 &0_2
    \end{pmatrix},
\end{equation}
which is simply a rearrangement of the modes.

One point to note is that we can pick $a$ and $b$ such that $\lambda_1=\lambda_2$ (specifically, by picking $a=b$). Despite the fact that, in this particular case, both sides of Eq.~(\ref{eq: s components}) go to $0$, the matrix $S$ that we have calculated for the general case is still correct.

\subsection{Three-mode example}
We will also show how to diagonalise a specific three-mode case. We consider covariance matrices of the form
\begin{equation}
    V=\begin{pmatrix}
    a \mathcal{I}_2 &c \mathcal{Z} &c \mathcal{Z}\\
    c \mathcal{Z} &a \mathcal{I}_2 &c \mathcal{I}_2\\
    c \mathcal{Z} &c \mathcal{I}_2 &a \mathcal{I}_2
    \end{pmatrix}.
\end{equation}
The corresponding symplectic eigenvalues are
\begin{align}
    &\lambda_1=a-c,\\
    &\lambda_2=\frac{\sqrt{2a^2+2ac-3c^2+\sqrt{x}}}{\sqrt{2}},\\
    &\lambda_2=\frac{\sqrt{2a^2+2ac-3c^2-\sqrt{x}}}{\sqrt{2}},
\end{align}
where
\begin{equation}
    x=4a^2+4ac-7c^2.
\end{equation}

Due to the greater number of quantities that must be calculated in the three-mode case, we will not present the explicit calculations here, but they are contained in the supplementary Mathematica file\cite{SUPP}. The diagonalising symplectic is given by
\begin{align}
    &S=\begin{pmatrix}
    0_2 &\frac{1}{\sqrt{2}}\mathcal{I}_2 &-\frac{1}{\sqrt{2}}\mathcal{I}_2\\
    \frac{2c\sqrt{\lambda_2}}{\sqrt{\sqrt{x}(ay_{+}+2\lambda_2^2)}}\mathcal{Z} &\frac{\sqrt{ay_{+}+2\lambda_2^2}}{2\sqrt{\lambda_2\sqrt{x}}}\mathcal{I}_2 &\frac{\sqrt{ay_{+}+2\lambda_2^2}}{2\sqrt{\lambda_2\sqrt{x}}}\mathcal{I}_2\\
    \frac{\sqrt{2a+y_{+}}}{\sqrt{2\sqrt{x}}}\mathcal{I}_2 &\frac{\sqrt{2a+y_{-}}}{2\sqrt{x}}\mathcal{Z} &\frac{\sqrt{2a+y_{-}}}{2\sqrt{x}}\mathcal{Z}
    \end{pmatrix},\\
    &y_{\pm}=c\pm\sqrt{x}.
\end{align}

The diagonalisation of a more general three-mode case, specifically one with a covariance matrix of the form
\begin{equation}
    V=\begin{pmatrix}
    a \mathcal{I}_2 &c_1 \mathcal{Z} &c_2 \mathcal{Z}\\
    c_1 \mathcal{Z} &b \mathcal{I}_2 &c_3 \mathcal{Z}\\
    c_2 \mathcal{Z} &c_3 \mathcal{Z} & \mathcal{I}_2
    \end{pmatrix},
\end{equation}
is also given in the Mathematica file. This is a very general class of state, with five free parameters, which contains the relevant states for idler-free channel position finding\cite{pereira_idler-free_2020}. Although the analytical expression for the resulting symplectic is complicated (and therefore not displayed here), it can be easily calculated using the method presented in this paper. It is computationally difficult to find a similar analytical form using existing methods.

\subsection{Degenerate three-mode example}\label{section: degenerate example}
By finding the diagonalising symplectic for a three-mode covariance matrix with a degenerate eigenvalue, we will demonstrate how degenerate symplectic eigenvalues can be handled. The covariance matrices that we consider take the form
\begin{equation}
    V^{(0)}=\begin{pmatrix}
    a \mathcal{I}_2 &\frac{a}{2} \mathcal{I}_2 &\frac{a}{2} \mathcal{I}_2\\
    \frac{a}{2} \mathcal{I}_2 &a \mathcal{I}_2 &\frac{a}{2} \mathcal{I}_2\\
    \frac{a}{2} \mathcal{I}_2 &\frac{a}{2} \mathcal{I}_2 &a \mathcal{I}_2
    \end{pmatrix},
\end{equation}
and have symplectic eigenvalues
\begin{equation}
    \lambda_1^{(0)}=2a,~~\lambda_2^{(0)}=\lambda_3^{(0)}=\frac{a}{2}.
\end{equation}
To deal with this problem, we introduce the perturbed covariance matrix
\begin{equation}
    \begin{split}
        V^{(\epsilon)}&=V^{(0)}+\epsilon\begin{pmatrix}
        \mathcal{I}_2 &0_2 &0_2\\
        0_2 &0_2 &0_2\\
        0_2 &0_2 &0_2
        \end{pmatrix}\\
        &=\begin{pmatrix}
        (a + \epsilon) \mathcal{I}_2 &\frac{a}{2} \mathcal{I}_2 &\frac{a}{2} \mathcal{I}_2\\
        \frac{a}{2} \mathcal{I}_2 &a \mathcal{I}_2 &\frac{a}{2} \mathcal{I}_2\\
        \frac{a}{2} \mathcal{I}_2 &\frac{a}{2} \mathcal{I}_2 &a \mathcal{I}_2
        \end{pmatrix},
    \end{split}
\end{equation}
which has symplectic eigenvalues
\begin{align}
    &\lambda_1^{(\epsilon)}=\frac{a}{2},\\
    &\lambda_2^{(\epsilon)}=\frac{\sqrt{17a^2+8a\epsilon+4\epsilon^2+\sqrt{x}(5a+2\epsilon)}}{2\sqrt{2}},\\
    &\lambda_3^{(\epsilon)}=\frac{\sqrt{17a^2+8a\epsilon+4\epsilon^2-\sqrt{x}(5a+2\epsilon)}}{2\sqrt{2}},
\end{align}
where
\begin{equation}
    x=9a^2-4a\epsilon+4\epsilon^2.
\end{equation}
These symplectic eigenvalues are non-degenerate for all $\epsilon>0$. It is clear from the form of $V^{(\epsilon)}$ that, if $V^{(0)}$ is a valid covariance matrix, $V^{(\epsilon)}$ is too, for all $x \geq 0$. The diagonalising symplectic for $V^{(\epsilon)}$ is given by
\begin{align}
    &S^{(\epsilon)}=\begin{pmatrix}
    0_2 &\frac{1}{\sqrt{2}}\mathcal{I}_2 &-\frac{1}{\sqrt{2}}\mathcal{I}_2\\
    \frac{y_{-}}{\sqrt{2}}\mathcal{I}_2 &\frac{y_{+}}{2}\mathcal{I}_2 &\frac{y_{+}}{2}\mathcal{I}_2\\
    -\frac{y_{+}}{\sqrt{2}}\mathcal{I}_2 &\frac{y_{-}}{2}\mathcal{I}_2 &\frac{y_{-}}{2}\mathcal{I}_2
    \end{pmatrix},\label{eq: non-deg S}\\
    &y_{\pm}=\frac{\sqrt{\sqrt{x}\pm (a-2\epsilon)}}{\sqrt[4]{x}}.
\end{align}
The explicit details of the calculation are again given in the supplementary Mathematica file. Now, since
\begin{equation}
    V^{(0)}=\lim_{\epsilon\to 0} V^{(\epsilon)},
\end{equation}
and since Eq.~(\ref{eq: non-deg S}) defines a diagonalising symplectic for any value of $\epsilon$, we can find the symplectic, $S^{(0)}$, that diagonalises the unperturbed covariance matrix, $V^{(0)}$, as the limit
\begin{equation}
    S^{(0)}=\lim_{\epsilon\to 0} S^{(\epsilon)}.
\end{equation}
We find that
\begin{equation}
    S^{(0)}=\begin{pmatrix}
    0_2 &\frac{1}{\sqrt{2}}\mathcal{I}_2 &-\frac{1}{\sqrt{2}}\mathcal{I}_2\\
    \frac{1}{\sqrt{3}}\mathcal{I}_2 &\frac{1}{\sqrt{3}}\mathcal{I}_2 &\frac{1}{\sqrt{3}}\mathcal{I}_2\\
    -\sqrt{\frac{2}{3}}\mathcal{I}_2 &\frac{1}{\sqrt{6}}\mathcal{I}_2 &\frac{1}{\sqrt{6}}\mathcal{I}_2
    \end{pmatrix}.
\end{equation}
This is still a valid symplectic matrix and it diagonalises $V^{(0)}$. Thus, we can find the diagonalising symplectic for a covariance matrix with degenerate eigenvalues by perturbing it such that it loses its degeneracy, diagonalising the perturbed covariance matrix and taking the limit as the magnitude of the perturbation goes to $0$.

\section{Relation with previous methods}\label{section: previous methods}

A common numerical algorithm for diagonalising covariance matrices 
involves the following steps \cite{PhysRevA.79.052327}. 

First we write 
\begin{equation}
	S = D^{-\frac{1}{2}}KV^{\frac{1}{2}},
\end{equation}
for a real, orthogonal transformation $K$. Such a matrix $S$ automatically verifies Eq.~(\ref{eq: v dec}) (recalling that the square root of a positive, symmetric matrix is also symmetric), yet in general it is not symplectic.

We now impose Eq.~(\ref{eq: s cond}), by writing
\begin{equation}
    V^{\frac{1}{2}} K^T D^{-\frac{1}{2}} \Omega D^{-\frac{1}{2}} K V^{\frac{1}{2}} = \Omega.
\end{equation}
Defining the anti-symmetric matrix
\begin{equation}
    X=V^{\frac{1}{2}}\Omega V^{\frac{1}{2}},
\end{equation}
we must find the unique, orthogonal matrix $K$ satisfying
\begin{equation}
    \begin{split}
        K X^{-1} K^T &= -D^{-\frac{1}{2}} \Omega D^{-\frac{1}{2}}\\
        &= \bigoplus_{m=1}^d -\frac{1}{\lambda_m}\omega,
    \end{split}
\end{equation}
where the $\lambda_m$ are the symplectic eigenvalues of $V$.
%to the canonical form \cite{zumino1962normal}.

We construct this $K$ as follows. Let $U$ be a unitary that diagonalises $X$ (and hence $X^{-1}$). Half of the eigenvectors of $X$ are the complex conjugates of the other half. We order the eigenvectors of $X$, $\mathbf{x}_m$, that compose $U$ as
\begin{equation}
    U=\begin{pmatrix}
    \mathbf{x}_1
    &\mathbf{x}_1^{*}
    &\mathbf{x}_2
    &\mathbf{x}_2^{*}
    &\ldots
    &\mathbf{x}_d
    &\mathbf{x}_d^{*}
    \end{pmatrix}.
\end{equation}
Then, $X^{-1}$ can be diagonalised as
\begin{equation}
    U^{\dagger}X^{-1}U=\bigoplus^d_{m=1} -\frac{i}{\lambda_m}\mathcal{Z}.
\end{equation}
We now define the matrix $\Gamma$ as
\begin{align}
    \Gamma=\bigoplus^d \gamma,~~
    \gamma=\frac{1}{\sqrt{2}}\begin{pmatrix}
    -i &i\\
    1 &1
    \end{pmatrix}.
\end{align}
Finally, we can write
\begin{equation}
    K=\Gamma U^{\dagger}.\label{eq: K expr}
\end{equation}
Since $\Gamma$ is unitary, $K$ is also a unitary. From the form of $\Gamma$ and Eq.~(\ref{eq: K expr}), we can see that $K^T$ is constructed as
\begin{equation}
    K^T=\sqrt{2}\begin{pmatrix}
    -\mathrm{Im}[\mathbf{x}_1]
    &\mathrm{Re}[\mathbf{x}_1]
    &\ldots
    &-\mathrm{Im}[\mathbf{x}_d]
    &\mathrm{Re}[\mathbf{x}_d]
    \end{pmatrix}.
\end{equation}
Consequently, $K$ is real, and therefore orthogonal.

We can therefore write
\begin{equation}
    \begin{split}
        K X^{-1} K^T &= K X^{-1} K^{\dagger}\\
        &= \Gamma U^{\dagger}X^{-1}U\Gamma^{\dagger}\\
        &= \bigoplus_{m=1}^d -\frac{i}{\lambda_m}\gamma\mathcal{Z}\gamma^{\dagger}\\
        &= \bigoplus_{m=1}^d -\frac{1}{\lambda_m}\omega,
    \end{split}
\end{equation}
showing that the symplectic condition is satisfied.
%The above algorithm is used in the strawberry fields library \cite{killoran2019strawberry}.

Alternative yet related algorithms for diagonalising quadratic bosonic Hamiltonians are also known in the nuclear physics community\cite{derezinski2017bosonic,blaizot1986quantum}. 

Compared to previous methods, our Theorem~\ref{theorem_main} offers the advantage of not requiring full diagonalisations. Whilst we need to find the determinants of certain submatrices, our technique does not involve taking matrix powers, and so does not require us to find the eigenvectors of matrices (only eigenvalues). This can often be simpler analytically. In the supplementary Mathematica file, we find that the algorithm presented in Section~\ref{section: main theory} is faster at finding the diagonalising symplectic for covariance matrices given by Eq.~(\ref{eq: CM 2-mode squeezed}) (parametrised by three free variables) than the existing method presented in this section, although the existing method is faster for numerical covariance matrices. This suggests that the method derived in this paper could be of use when the analytical form of the diagonalising symplectic for a covariance matrix with free variables is required.

\hfill
\section{Conclusion}

In this paper, we have presented a simple technique for finding the diagonalising symplectic for a covariance matrix, based on similar ideas to the technique for calculating the diagonalising unitary for a Hermitian matrix presented by Denton et al.\cite{denton_eigenvectors_2021} This technique requires the calculation of submatrix determinants, but does not require eigenvector decompositions (which can often be analytically difficult). Our technique applies to all covariance matrices that do not have degenerate symplectic eigenvalues, but we have also presented a workaround for matrices that have degenerate symplectic eigenvalues. The method can also be extended to certain indefinite and non-real, symmetric matrices. We have provided three example cases, to show how our method can be applied.

Whilst methods to find the diagonalising symplectic for a covariance matrix already exist, the technique presented here may provide a simpler technique for calculating the diagonalising symplectic, in some cases.

\smallskip
\textbf{Acknowledgments.}~This work was funded by the European Union's Horizon 2020 Research and Innovation Action under grant agreement No. 862644 (FET-OPEN project: Quantum readout techniques and technologies, QUARTET). L.B.~acknowledges support by the program ``Rita Levi  Montalcini'' for
young researchers.

\end{document}